\documentstyle[12pt,aaspp4,flushrt]{article}
\input epsf

\newcommand\g{\gamma}

\newcommand\s{\sigma}
\newcommand\e{\varepsilon}
\renewcommand\t{\theta}
\newcommand\tg{\t_\g}
\newcommand\cg{c_\g}
\newcommand\rc{r_{\rm crit}}
\newcommand\uc{u_{\rm caus}}
\newcommand\vc{v_{\rm caus}}
\newcommand\M{{\cal M}}
\newcommand\bSIS{b_{\rm SIS}}

\newcommand\x{{\vec x}}
\newcommand\uu{{\vec u}}

\newcommand\refeq[1]{eq.~(\ref{eq:#1})}

\newcommand\refeqs[2]{eqs.~(\ref{eq:#1}) and (\ref{eq:#2})}

\begin{document}

\title{Gravitational Lenses With More Than Four Images: \\
  I. Classification of Caustics}
\author{Charles R. Keeton$^{1}$, Shude Mao$^{2,3}$,
  and Hans J. Witt$^{4}$}
\affil{$^{1}$ Steward Observatory, University of Arizona, Tucson, AZ 85721, USA}
\affil{$^{2}$ Univ. of Manchester, Jodrell Bank Observatory,
  Macclesfield, Cheshire SK11 9DL, UK}
\affil{$^{3}$ Max-Planck-Institut f\"ur Astrophysik,
  Karl-Schwarzschild-Strasse 1, 85740 Garching, Germany}
\affil{$^{4}$ Astrophysikalisches Institut Potsdam, An der Sternwarte 16,
  14482 Potsdam, Germany}

\bigskip
\bigskip
\centerline{Accepted for publication in {\it The Astrophysical Journal\/}}

\begin{abstract}
We study the problem of gravitational lensing by an isothermal
elliptical density galaxy in the presence of a tidal perturbation.
When the perturbation is fairly strong and oriented near the
galaxy's minor axis, the lens can produce image configurations with
six or even eight highly magnified images lying approximately on a
circle. We classify the caustic structures in the model and
identify the range of models that can produce such lenses. Sextuple
and octuple lenses are likely to be rare because they require
special lens configurations, but a full calculation of the
likelihood will have to include both the existence of lenses with
multiple lens galaxies and the strong magnification bias that
affects sextuple and octuple lenses. At optical wavelengths these
lenses would probably appear as partial or complete Einstein rings,
but at radio wavelengths the individual images could probably be
resolved.
\end{abstract}

\section{Introduction}

The first gravitational lens to be discovered, Q~0957+561, has a
simple double image configuration (Walsh, Carswell \& Weymann
1979). It was quickly followed by the first four-image lens,
PG~1115+080 (Weymann et al.\ 1980). Together with Einstein ring
lenses produced by extended sources (e.g.\ MG~1131+0456, Hewitt et
al.\ 1988), double and quadruple lenses nearly exhaust the list of
configurations in the more than 50 known strong gravitational
lenses.\footnote{For a summary, see
http://cfa-www.harvard.edu/glensdata.} The only exceptions are
2016+112, which has three images and appears to require the rare
and complicated situation of two lens galaxies at different
redshifts (Lawrence et al.\ 1984; Nair \& Garrett 1997), and
B~1933+503, which has ten images that can be nicely explained as a
combination of three distinct sources, two of which are
quadruply-imaged and one of which is doubly-imaged (Sykes et al.\
1998).

Theoretical studies have shown, though, that lenses with more than
four images of a single source can exist. Schneider, Ehlers \&
Falco (1992) gave a mathematical analysis of caustic structures that
yield additional images. Kochanek \& Apostolakis (1988) surveyed
models with two spherical galaxies at different redshifts and found
that they can produce up to seven images. Witt \& Mao (2000) found
examples of models with an elliptical density galaxy and an external
shear that can produce up to eight images, but they did not do a
full survey of the models. Lenses with more than four images would
be not only interesting to observe, but also very useful in
determining the lensing mass distribution using the numerous
constraints.

In this paper we present a systematic classification of the
caustics and image configurations for lenses consisting of an
isothermal elliptical density galaxy and a tidal perturbation. We
show examples of lensing cross sections and magnification
distributions for different image numbers. In a forthcoming paper
we will study the observability of lenses with double, quadruple,
sextuple, and octuple image configurations. The outline of this
paper is as follows. In \S 2 we review basic lens theory and the
singular isothermal ellipsoid (SIE) and external shear lens models.
In \S 3 we present an analytic classification of the caustics in a
lens model with an SIE galaxy and a tidal perturbation approximated
as an external shear. In \S 4 we study the stability of the
caustics by adding a small core radius to the galaxy and letting
the perturbation be produced by a neighboring galaxy or group.
Finally, in \S 5 we discuss some of the observational consequences
and applications of our results.

\section{Methods}

\subsection{Basic lens theory}

The theory of gravitational lensing is discussed in detail by
Schneider et al.\ (1992); here we summarize the features central
to our analysis. The mapping between a source at angular position
$\uu$ on the sky and an image at angular position $\x$ is given
by the lens equation,
\begin{equation}
  \uu = \x - \nabla\phi(\x)\,, \label{eq:lens}
\end{equation}
where the lensing potential $\phi$ is the projected gravitational
potential of the lens mass. The potential is determined by the
two-dimensional Poisson equation $\nabla^2\phi(\x) =
2\Sigma(\x)/\Sigma_{\rm cr}$, where $\Sigma$ is the projected
surface mass distribution of the lens and the critical surface
density for lensing is
\begin{equation}
  \Sigma_{\rm cr} = {c^2 \over 4 \pi G}\,{D_{\rm os} \over
    D_{\rm ol} D_{\rm ls}}\ , \label{eq:sigcr}
\end{equation}
where $D_{\rm ol}$ and $D_{\rm os}$ are angular diameter distances
from the observer to the lens and source, respectively, and
$D_{\rm ls}$ is the angular diameter distance from the lens to the
source. For a point source at $\uu$, there is an image at each root
$\x_i$ of the lens equation (\ref{eq:lens}).

The brightnesses of the images are determined by the magnification
tensor
\begin{equation}
  \M(\x) \equiv \left({\partial\uu \over \partial \x}\right)^{-1}
  = \left[\matrix{
    1-\phi_{,xx}(\x) &  -\phi_{,xy}(\x) \cr
     -\phi_{,xy}(\x) & 1-\phi_{,yy}(\x) \cr
  }\right]^{-1} , \label{eq:mu}
\end{equation}
where subscripts denote partial differentiation, $\phi_{,ij} =
\partial^2\phi / \partial x_i \partial x_j$. The magnification of a
point image at position $\x$ is given by $|\det\M(\x)|$. In general there
are several curves in the image plane along which the magnification
tensor is singular and the magnification is infinite ($\det\M^{-1}
= 0$). These are called ``critical curves,'' and they map to
``caustics'' in the source plane. Caustics mark discontinuities in
the number of images, which leads to the key idea for our analysis:
in order to determine the number of images produced by a lens
model, it is sufficient to examine the caustics. If we know that a
source arbitrarily far from the lens produces only one image, then
we can imagine moving the source around and noting when it crosses
a caustic to keep track of the total number of images. We discuss
examples in \S 3.2.

\subsection{Model components}

We model a galaxy as a singular isothermal ellipsoid, because this
model is not only analytically tractable but also consistent with
models of individual lenses, lens statistics, stellar dynamics,
and X-ray galaxies (e.g.\ Fabbiano 1989; Maoz \& Rix 1993; Kochanek
1995, 1996; Grogin \& Narayan 1996; Rix et al.\ 1997). If we take
the ellipsoid to be oblate with intrinsic axis ratio $q_3$ then
its three-dimensional density distribution is
\begin{equation}
  \rho = {\s^2 \over 2\pi G q_3}\,{\e_3 \over \sin^{-1}\e_3}\,
    {1 \over R^2 + z^2/q_3^2}\,,
\end{equation}
where $G$ is the gravitational constant, $\s$ is the velocity
dispersion and $\e_3 = \sqrt{1-q_3^2}$ the eccentricity of the mass
distribution, and $(R, z)$ are usual cylindrical coordinates. For
lensing we need the projected mass distribution. Choosing
coordinates with the projected major axis along the $x$-axis, the
projected surface density distribution is
\begin{eqnarray}
  {\Sigma \over \Sigma_{\rm cr}} &=& {b_I \over 2 q}\,
    {1 \over \sqrt{x^2+y^2/q^2}}\ , \\
  \mbox{where}\quad
  b_I &=& 4\pi\,{\e_3 \over \sin^{-1}\e_3}\,\left({\s \over
  c}\right)^2\,{D_{\rm ls} \over D_{\rm os}}\ . \label{eq:bI}
\end{eqnarray}
The projected axis ratio $q$ depends on the intrinsic axis ratio
and the inclination angle $i$ (where $i=0^\circ$ is face-on and
$i=90^\circ$ is edge-on),
\begin{equation}
  q = \sqrt{q_3^2\,\sin^2 i + \cos^2 i}\,.
\end{equation}
The lensing properties of the isothermal ellipsoid have been given
by Kassiola \& Kovner (1993), Kormann, Schneider \& Bartelmann
(1994), and Keeton \& Kochanek (1998). The lensing potential $\phi$
and the deflection angle $(\phi_{,x},\phi_{,y})$ are
\begin{eqnarray}
  \phi &=& x\,\phi_{,x} + y\,\phi_{,y}\,, \label{eq:SIE}\\
  \phi_{,x} &=& {b_I \over \sqrt{1-q^2}}\,\tan ^{-1}\left(\sqrt{1-q^2 \over q^2 x^2 + y^2}\,x\right)\,, \nonumber\\
  \phi_{,y} &=& {b_I \over \sqrt{1-q^2}}\,\tanh^{-1}\left(\sqrt{1-q^2 \over q^2 x^2 + y^2}\,y\right)\,.
    \nonumber
\end{eqnarray}

In the limit of a spherically symmetric mass distribution (a
singular isothermal sphere, or SIS), the tangential critical curve
is a circle with radius
\begin{equation}
  \bSIS = 4\pi\,\left({\s \over c}\right)^2\,{D_{\rm ls}
    \over D_{\rm os}}
  \label{eq:bSIS}
\end{equation}
(in angular units); the maximum separation between images is $\approx
2\bSIS$. This ``critical radius'' therefore serves as a natural
length scale in the lensing analysis, and we use it this way below.
Typically $\bSIS \sim 0.2\arcsec$--$3\arcsec$ for galaxy-scale
lenses.

In \S 3 we study models with a tidal perturbation approximated as
an external shear. The potential and deflection angle for an
external shear are
\begin{eqnarray}
  \phi &=& -{1\over2} \g r^2 \cos 2(\t-\tg)\,, \label{eq:shear} \\
  \phi_{,x} &=&          -  \g r \cos(\t-2\tg) = -\g x \cos 2\tg - \g y \sin 2\tg \nonumber\,, \\
  \phi_{,y} &=& \phantom{-} \g r \sin(\t-2\tg) = -\g x \sin 2\tg + \g y \cos 2\tg \nonumber\,,
\end{eqnarray}
where $\g$ is the strength of the shear and $\tg$ is its direction
angle, which is equal to the angle between the major axis of the
galaxy and the shear axis because we have aligned the galaxy with
the $x$-axis. Note that a shear described by $(\g,\tg)$ is
equivalent to one described by $(-\g,\tg+\pi/2)$; we avoid this
ambiguity by considering only shears with $\g>0$ to be physical.

\section{Galaxy+Shear Models}

A common lens system features a galaxy that is well modeled as an
isothermal ellipsoid, plus a perturbation from objects at the same
redshift as the lens galaxy (e.g.\ neighboring galaxies, or a group
or cluster) or objects along the line of sight. For examples, see
Hogg \& Blandford (1994), Keeton, Kochanek \& Seljak (1997),
Kundi\'c et al.\ (1997ab), Witt \& Mao (1997), Tonry (1998), and
Tonry \& Kochanek (1999). To lowest order, the perturbation can be
approximated as an external shear as in \refeq{shear}. We begin by
studying such galaxy+shear lens models because they admit a
complete analytic treatment. In \S 4.2 we consider models in which
the perturbation is instead produced by a discrete object like a
galaxy or group.

\subsection{Critical curves and caustics}

To obtain a galaxy+shear lens model we merely add the lensing
potentials in \refeqs{SIE}{shear}. The magnification from the joint
model has a simple analytic form,
\begin{eqnarray}
  \det\M^{-1} = 1 - \g^2 - \sqrt{2}\,{b_I \over r}\,
    {1 + \g \cos 2(\t-\tg) \over \sqrt{ (1+q^2) - (1-q^2)\cos 2\t }}\,.
    \label{eq:mag}
\end{eqnarray}
This lens model has a single critical curve, or curve of infinite
magnification. From \refeq{mag} we can easily find a polar
parametric form for this curve,
\begin{equation}
  \rc(\t) = {\sqrt{2}\, b_I \over 1-\g^2}\,
    {1 + \g \cos 2(\t-\tg) \over \sqrt{ (1+q^2) - (1-q^2)\cos 2\t }}\,.
    \label{eq:crit}
\end{equation}
The corresponding caustic is found by mapping the critical curve
to the source plane with the lens equation (\ref{eq:lens}).  The
caustic can be written in a cartesian parametric form,
\begin{eqnarray}
  \uc(\t) &=& \left[ \cos\t + \g \cos(\t-2\tg) \right] \rc(\t)
    - {b_I \over \sqrt{1-q^2}}\,\tan^{-1}\left(\xi\,\cos\t\right)\,,
    \label{eq:astr}\\
  \vc(\t) &=& \left[ \sin\t - \g \sin(\t-2\tg) \right] \rc(\t)
    - {b_I \over \sqrt{1-q^2}}\,\tanh^{-1}\left(\xi\,\sin\t\right)\,,
    \nonumber\\
  \mbox{where}\quad
    \xi &=& \left[ 2(1-q^2) \over (1+q^2) - (1-q^2)\cos 2\t \right]^{1/2} .
    \nonumber
\end{eqnarray}
Note that in \refeq{crit} the parameter $\t$ is interpreted as the
polar angle, but in \refeq{astr} $\t$ is simply a parameter that
runs from 0 to $2\pi$.

The caustic is continuous but not smooth; in general it has four or
more cusps. (See examples in \S 3.2.) We can find a simple equation
that gives the location of the cusps. Consider the parametric
derivatives of the caustic,
\begin{eqnarray}
  \left[ \matrix{ d\uc/d\t \cr d\vc/d\t } \right]
  &=& - b_I\,\xi\,{ 3\g\sin2(\t-\tg) +
    (1+\g\cos2(\t-\tg))\xi^2\sin\t\cos\t \over (1-\g^2)\sqrt{1-q^2} }
    \nonumber\\
  && \quad \times \left[ \matrix{ \cos\t + \g\cos(\t-2\tg) \cr
    \sin\t - \g\sin(\t-2\tg) } \right] . \label{eq:astrderiv}
\end{eqnarray}
At a cusp the caustic stops and changes direction, so both
$d\uc/d\t$ and $d\vc/d\t$ vanish (and at least one of them changes
sign). The only way for both derivatives to vanish simultaneously
is for the common multiplicative factor in \refeq{astrderiv} to
vanish. We can simplify this to find that the condition for a cusp
is
\begin{equation}
  \g \sin 2(\t-\tg) \Bigl[ 3(1+q^2) - 2(1-q^2) \cos 2\t \Bigr]
  + (1-q^2)(\sin 2\t + \g \sin 2\tg) = 0\,. \label{eq:cusps}
\end{equation}

If our model galaxy were non-singular, the lens model would have a
second critical curve and caustic. Since it is singular, however,
the second critical curve collapses to a point at the origin, and
the corresponding caustic is considered to be a ``pseudo-caustic.''
The pseudo-caustic has the cartesian parametric form
\begin{eqnarray}
  u_{\rm pseudo}(\t) &=&
    - {b_I \over \sqrt{1-q^2}}\,\tan^{-1}\left(\xi\,\cos\t\right)\,,
    \label{eq:pseudo} \\
  v_{\rm pseudo}(\t) &=&
    - {b_I \over \sqrt{1-q^2}}\,\tanh^{-1}\left(\xi\,\sin\t\right)\,,
    \nonumber
\end{eqnarray}
where again $0 \le \t \le 2\pi$, and $\xi$ is the same as in
\refeq{astr}. Note that the pseudo-caustic does not depend on the
shear.

\subsection{Examples}

Figure 1 shows typical caustics and pseudo-caustics for
galaxy+shear lens models, plotted using $\bSIS$ from \refeq{bSIS}
as the natural length scale. In general the pseudo-caustic is
smooth, while the true caustic has cusps that give it a diamond or
astroid shape. As discussed in \S 2.1, the caustics indicate where
the number of images changes. A source outside both the caustic and
the pseudo-caustic produces one image. When the source crosses to
the inside of the pseudo-caustic it gains one additional image,
which is faint, close to the galaxy center, and distorted radially
relative to the galaxy; the pseudo-caustic is sometimes labeled the
``radial caustic'' because of the nature of the distortions. When
the source crosses to the inside of the astroid caustic it gains
two more images, which are bright, close to the corresponding
critical curve, and distorted tangentially relative to the galaxy;
hence this caustic is sometimes labeled the ``tangential caustic.''
The presence of the pseudo-caustic and the tangential caustic means
that the number of images in these models is either one, two, or
four depending on the location of the source. All of these features
are generic to singular lens models that are not axisymmetric (see
Schneider et al.\ 1992).

The size and shape of the astroid caustic are determined by the net
quadrupole moment of the lens model. When the galaxy and shear are
aligned ($\tg = 0^\circ$, e.g.\ Figure 1a), the individual
quadrupoles from the galaxy ellipticity and the shear combine to
produce a larger astroid caustic than produced by either one alone.
In other words, the area where sources produce 4-image lenses is
larger. Mild misalignment between the ellipticity and shear twists
the caustic (e.g.\ Figure 1b). As the shear becomes more misaligned
(e.g.\ Figure 1c), the quadupole from the shear partially cancels
the quadrupole from the ellipticity; this is why Figure 1c has the
smallest astroid caustic despite having the largest shear. We
return to this point in the discussion (\S 5).

Figure 2 shows that when the shear is nearly orthogonal to the
ellipticity ($\tg \simeq 90^\circ$), it can nearly cancel the
effects of the ellipticity to produce a caustic that is quite
small. Nevertheless, the caustic is very interesting because it
shows qualitatively new features not seen in Figure 1. (Similar
examples can be found in Witt \& Mao 2000.) The caustic folds over
on itself in features called ``swallowtails'' (see Schneider et
al.\ 1992). Because the number of images increases by two when the
source crosses the caustic, a source inside a swallowtail produces
six images (e.g.\ Figure 2b). The swallowtails are sensitive to the
angle between the ellipticity and shear. They grow larger as the
misalignment increases, until with near perfect misalignment the
swallowtails overlap (e.g.\ Figure 2c). The region of the source
plane where swallowtails overlap corresponds to sources that
produce eight images. In other words, the interaction of the
ellipticity and shear makes it possible to have new image
configurations with six or even eight images.

Not all combinations of ellipticity and shear can produce these new
image configurations. Figure 3 shows caustics for three cases where
the shear is orthogonal to the ellipticity ($\tg=90^\circ$). If the
shear is small (Figure 3a), the system is dominated by the
ellipticity and any swallowtails that exist are small. If the shear
is large (Figure 3c), the system is dominated by the shear and
again any swallowtails are small. For a given ellipticity, only a
narrow range of orthogonal shears can produce overlapping
swallowtails (Figure 3b). In \S 3.3 we discuss in detail the range
of models that can produce 6- and 8-image configurations.

\subsection{Models that can produce more than 4 images}

Appendices A and B give a mathematical analysis of the range of
models that have swallowtails and thus can produce 6- or 8-image
lenses; we summarize the results here. We use the presence of
swallowtails in the caustic to indicate that a model can produce
more than four images. This approach does not directly indicate
the probability of observing a lens with more than four images.
Estimating this probability requires detailed computations of
lensing cross sections and magnification distributions for
realistic sets of lens environments. We discuss these issues
briefly in \S 5 and plan to study them in more detail in a
forthcoming paper. For now we seek to delineate the conditions
under which a lens can produce more than four images.

Figures 4--6 show the envelope of swallowtail models in the
$(q,\g)$ plane for different values of the shear angle $\tg$. When
the shear is orthogonal to the ellipticity (Figure 4), the envelope
encloses small shears for modestly flattened galaxies ($q \lesssim
1$) and larger shears for more flattened galaxies ($q \ll 1$).
Swallowtails can exist for $\g$ arbitrarily small and $q$
arbitrarily close to unity (ellipticity arbitrarily small),
provided that the shear and ellipticity are in the narrow band
where they properly balance each other. A relatively broad band of
the swallowtail models also have overlapping swallowtails and thus
can produce 8-image lenses.

If $\tg$ changes by even a few degrees away from perfect
misalignment, however, the envelope pinches off at the
high-$q$/low-$\g$ end (Figure 5). In other words, obtaining
swallowtails requires more ellipticity and shear. Also, the range
of models with overlapping swallowtails quickly disappears (not
shown). From these results we conclude that 6- and 8-image lenses
produced by small tidal perturbations (e.g.\ $\g \sim 0.1$) are
likely to be very rare because they require a special combination
of galaxy axis ratio and shear misalignment angle. However, with
stronger perturbations the range of swallowtail models is
considerably larger. Thus we predict that even though 6- or 8-image
lenses may still be rare, they are more likely to occur when the
perturbation is strong (such as a second galaxy near the primary
lens galaxy) than when the perturbation is weak. Since strong
perturbations may not be well approximated by the external shear
model, in \S 4.2 we study models using a full treatment of a
perturbation from a nearby galaxy or group.

The mathematical analysis does reveal that swallowtails can exist
for smaller shear misalignments; while the results are formally
interesting, they are physically implausible because they require
highly flattened galaxies and very large shears (Figure 6). For
$\tg \sim 60^\circ$, there are two envelopes that are comparable in
size.\footnote{The second envelope exists even for larger
misalignments, as indicated by Figure 12 in Appendix A. However, it
is too small to be seen in Figure 5.} For $45^\circ < \tg <
50.45^\circ$ ($0 > \cos2\tg > -1/\sqrt{28}$, see eq.~\ref{eq:a6}),
one of the envelopes stretches to arbitrarily large shears. For
$39.55^\circ < \tg < 45^\circ$ ($1/\sqrt{28} > \cos2\tg > 0$), the
finite envelope disappears but the infinite one remains. This
envelope finally disappears for $\tg < 39.55^\circ$ ($\cos2\tg >
1/\sqrt{28}$). With the strong perturbations required by these
envelopes, however, the simple shear approximation almost certainly
breaks down. Thus this analysis is probably not fully valid, but it
does suggest features to look for when studying models with very
strong perturbations, such as interacting galaxies.

\section{Stability of the Caustics}

\subsection{Non-singular lens models}

Although we have studied singular lens models for analytic
convenience, real galaxies may have a small but finite core radius.
It is important to understand whether swallowtails are robust under
the addition of a core radius. Adding a core radius adds one faint
image near the center of the galaxy for any configuration with more
than one image, so the total number of images is always odd (see
Schneider et al.\ 1992). Figure 7 shows generalizations of the
caustics in Figures 2b and 2c to a finite core radius $s$, using
the lensing properties of a softened isothermal ellipsoid given by
Keeton \& Kochanek (1998). The swallowtails shrink as $s$ increases
-- probably because when we increase $s$ with $\bSIS$ held fixed,
we decrease the galaxy mass and hence reduce the contribution of
the ellipticity to the potential. However, no convincing case of a
faint central image has been observed, and this limits the core
radius ($s/\bSIS \lesssim 0.1$, see Wallington \& Narayan 1993;
Kochanek 1996) and suggests that lens galaxies are quite cuspy. The
presence of such small cores would not significantly affect our
results.

\subsection{Two-galaxy models}

As discussed in \S 3.3, the galaxy+shear models suggest that the
perturbation must be relatively strong in order to produce
swallowtails. For such strong perturbations, the external shear
approximation may not be justified, so in this section we examine a
simple model in which the perturbation is produced by a second mass
distribution representing a neighboring galaxy or group. We again
use a singular isothermal mass distribution for the perturber, but
for simplicity we assume it is spherical.

The perturber is described by three physical parameters: its
velocity dispersion $\s_2$ (or equivalently its critical radius
$b_{\rm SIS,2}$ given by eq.~\ref{eq:bSIS}),\footnote{In this
section we use a subscript 1 to denote the main lens galaxy and a
subscript 2 to denote the perturber. We continue to use the
critical radius $b_{\rm SIS,1}$ of the main lens galaxy as the
natural scale length.} and its position relative to the lens
galaxy, given by the projected distance $d$ from the galaxy center
and the angle $\tg$ from the lens galaxy's major axis. To
generalize the galaxy+shear models, it is convenient to
characterize the perturber not by its velocity dispersion $\s_2$
but rather by the strength of the perturbation. To lowest order,
the perturbation is equivalent to an external shear with strength
\begin{equation}
  \g = {b_{\rm SIS,2}/(2d) \over 1 - b_{\rm SIS,2}/(2d)}\ .
\end{equation}
It is important to understand what this strength means. All
perturbers with a given strength $\g$ are equivalent to each other
and to an external shear of strength $\g$ -- {\it to second order
in the potential\/}. The differences enter only in terms of third
order and higher. Nevertheless, we show here that the differences
are important.

We note that since the perturbation strength $\g$ is a combination
of the velocity dispersion and distance, perturbers with a given
strength but different distances must have different velocity
dispersions,
\begin{equation}
  \left({\s_2 \over \s_1}\right)^2 = {b_{\rm SIS,2} \over b_{\rm SIS,1}}
  = 2\,{d \over b_{\rm SIS,1}}\,{\g \over 1+\g}\ . \label{eq:sigrat}
\end{equation}
In other words, the more distant the perturber is, the more massive
it has to be, and vice versa.

The top panels in Figure 8 show the caustics for a galaxy+shear
model with $q=0.5$ and $\tg=90^\circ$. They are similar to Figure 3
but include more values of $\g$. The other panels in Figure 8 show
the generalization to two-galaxy models, using the same values of
$\g$ and distances $d/b_{\rm SIS,1} = 5, 10, 20, 40$. Note that the
galaxy+shear models in the top panel are equivalent to two-galaxy
models with $d \to \infty$. Figure 9 is similar, but shows the
caustic structures for $\tg=88^\circ$. From \refeq{sigrat}, the
models in Figures 8 and 9 have perturbers that range in mass from
the scale of a galaxy to that of a cluster.

The examples show two important effects. First, the swallowtails
tend to be {\it larger\/} in two-galaxy models than in equivalent
galaxy+shear models, and they grow as the distance $d$ decreases.
The various models differ only in higher order terms, but those
terms apparently strengthen the perturber's effects. Equivalently,
the range of models that produce swallowtails is larger. Second, in
galaxy+shear models the symmetry of the shear implies that there
are always two identical swallowtails. By contrast, in two-galaxy
models one of the swallowtails is often quite large, while the
other is either small or absent. As a result, the area in the
source plane where swallowtails overlap is small or absent, so the
models have a limited ability to produce 8-image lenses.

We conclude from these examples that the qualitative features we
saw in the galaxy+shear models are stable when we change the source
of the perturbation. In fact, the range of swallowtail models seems
to {\it increase\/} with more realistic treatments of strong or
close perturbations. However, it appears that the overlapping
swallowtails required to produce octuple lenses are rare and not
very stable.

\section{Summary and Discussion}

We have studied the lensing properties of an isothermal elliptical
density galaxy in the presence of a tidal perturbation to classify
the caustic structures and identify different image configurations.
In most cases the models have a simple caustic structure
corresponding to standard 2-image and 4-image lens configurations.
However, when the tidal shear has a magnitude and direction
appropriate to partially cancel the effects of the galaxy's
ellipticity, the caustics develop complicated swallowtail features
that correspond to 6- and 8-image configurations that have not yet
been observed. We gave a complete analytic treatment of the case of
a singular galaxy with a perturbation modeled as an external shear,
but we showed that the caustic structures are stable when one adds
a small core radius or uses a more realistic treatment of the
perturbation. In fact, the swallowtail caustic structures are
generally bigger with the more realistic perturbation than with the
shear approximation, which may enhance the likelihood of observing
lenses with more than four images.

Our analysis has several observational consequences. First,
sextuple and octuple lenses are likely to be rare because they
require special lens configurations. In fact, they will probably be
found only when there is a second galaxy close enough to the lens
galaxy to provide a strong perturbation. While this situation is
uncommon, it is not exceedingly rare: at least five lenses\footnote{
B~1127+385 (Koopmans et al.\ 1999), B~1359+154 (Rusin et al.\ 1999),
B~1608+656 (Koopmans \& Fassnacht 1999), 2016+112 (see Nair \&
Garrett 1997 and references therein), and B~2114+022 (Augusto,
Wilkinson \& Browne 1996; Jackson et al.\ 1998).} appear to have
multiple lens galaxies within the Einstein radius. Still, to
quantify the likelihood of finding a sextuple or octuple lens,
it is necessary to compute the lensing cross sections for various
combinations of the galaxy axis ratio $q$ and the shear amplitude
$\g$ and misalignment angle $\tg$. Figure 10 shows cross sections
for $q=0.5$, $\g=0.22$, and three values of $\tg$ that correspond
to the cases shown in Figure 2. The configurations with more images
have smaller cross sections and thus smaller probabilities for
being observed. However they also have significantly higher
magnifications, which means that magnification bias will be
important (e.g.\ Turner 1980; Turner, Ostriker \& Gott 1984).
Magnification bias will mitigate the effects of small cross
sections to increase the likelihood of observing a sextuple or
octuple lens. Clearly a realistic prediction of the lensing
probabilities will require computation of cross sections and
magnification bias for many combinations of $(q,\g,\tg)$ weighted
by realistic populations of lens galaxies and perturbers. We
plan to address these issues of observability in a forthcoming
paper, both in the context of current surveys and of the Next
Generation Space Telescope, where large numbers of lenses are
expected (e.g.\ Barkana \& Loeb 1999).

Second, if discovered the sextuple and octuple lenses that we have
described will be easy to identify because the lensed images lie
approximately on a circle (see Figure 2). Any lens with more than
four images that do not trace a circle must not be of the type
described here. Indeed, in B~1933+503 there are ten images not on a
circle, and it is thought that they must be associated with three
different sources (Sykes et al.\ 1998). In B~1359+154 there are six
radio sources with four sources in a standard quadruple lens
configuration, plus two sources inside the configuration whose
interpretation is not clear (Myers et al.\ 1999). Although
observations and models suggest that there may be multiple lens
galaxies (Rusin et al.\ 1999), the fact that the six sources do not
follow a circle suggests that lens cannot be explained by
swallowtails produced by the lens galaxies.

Third, observed sextuple or octuple lenses would be very useful for
constraining models of the lensing mass distribution. The ``extra''
images beyond the standard two or four would provide additional
position and flux constraints. Even better, the sensitivity of the
caustic structures to the lens galaxy ellipticity and the shear
amplitude and misalignment angle means that the mere existence of
six or more images should place strong constraints on those
properties of the model. As a result, a sextuple or octuple lens
could represent a wonderful ability to break the common degeneracy
between the lens galaxy shape and the shear from the surrounding
environment (see Keeton et al.\ 2000a). Such a lens might even
serve as the long-sought ``golden lens'' for measuring the Hubble
constant $H_0$ (see Schechter 2000) provided that a time delay
could be measured, although the time delay might be relatively
short (a few days) because the images are all close to the critical
curve.

These applications are predicated on the ability to resolve the
individual images, and this ability might be limited when we
include the effects of finite source size. Figure 11 shows that the
source does not have to be very large before the images smear into
a partial or complete Einstein ring. This happens for $R_{\rm
src}/\bSIS \sim 0.01$, where typically $\bSIS \sim
1\arcsec$--$3\arcsec$ for galaxy-scale lenses. Radio surveys can
achieve sub-milli-arcsecond resolution with VLBI or VLBA mapping
(e.g.\ the CLASS survey, see Browne 2000 and references therein),
so they should still be able to resolve individual images and thus
find lenses amenable to these applications. By contrast, optical
surveys (such as the Sloan Digital Sky Survey, see Gunn et al.\
1998 and Fischer et al.\ 1999, or the Next Generation Space
Telescope, see Barkana \& Loeb 1999) are more likely find partial
or complete Einstein rings. Still, new techniques for modeling
Einstein rings show that they are very useful for constraining not
only the lens model but also the intrinsic shape of the source
(Keeton, Kochanek \& McLeod 2000b).

Finally, our study may be relevant for the so-called ellipticity
``crisis,'' where lens galaxies are inferred to have larger
ellipticities than the observed early-type galaxies (see Kochanek
1996 and references therein). From Figure 1, it is clear that the
caustic structures are the largest when the shear is aligned with
the major axis of the lensing galaxy so the ellipticity and shear
act coherently; conversely, the caustic structures are the smallest
when the shear is orthogonal to the major axis since the shear
partially cancels the lens ellipticity. Observationally, this means
that for a sample of quadruple lenses the shear may preferentially
lie along the galaxy major axis; a good example is B~1422+231 (Hogg
\& Blandford 1994). The presence of an aligned shear is difficult
to infer from lens models due to a degeneracy in the lens equation
(see Witt 1996): naive lens models simply imply a model ellipticity
equivalent to the combination of the true ellipticity and the
shear. This effect may generate a bias toward larger inferred
ellipticities for lens galaxies, although a full account of this
bias awaits further investigations.

\acknowledgements
{\it Acknowledgements.\/} We thank Peter Schneider for helpful
discussions, and the anonymous referee for prompt and helpful
comments that improved the discussion.

\appendix

\section{Models that can produce at least 6 images}

In this Appendix we compute the range of galaxy+shear models (see
\S 3) that can produce at least six images. Because the caustics
determine the image number, finding models that can yield at least
six images is equivalent to finding models in which the caustic has
swallowtails. We saw in \S 3.2 that in models without swallowtails
the caustic has four cusps, while in models with swallowtails the
caustic has more than four cusps. Thus to identify models with
swallowtails it is sufficient to find models that have more than
the standard four solutions to the cusp equation (\ref{eq:cusps}).

In other words, the envelope bounding the region in parameter space
where models have swallowtails is located where the cusp equation
develops additional pairs of solutions. If the cusp equation
(\ref{eq:cusps}) is written as $f(\t)=0$, the place where
additional solutions appear is defined by
\begin{equation}
  f(\t) = 0 \qquad\mbox{and}\qquad
  {\partial f(\t) \over \partial \t} = 0\,.
  \end{equation}
These equations yield two polynomials in $\sin 2\t$ and $\cos 2\t$,
so we can use the resultant method (e.g.\ Walker 1955; Erdl \&
Schneider 1993) to eliminate $\t$. We find that the envelope is
given by roots of the equation
\begin{equation}
  F(\g,\tg,q) \equiv \sum_{i=0}^6 a_i(\tg,q)\, (1-q^2)^{6-i}\, \g^i = 0\,,
  \label{eq:F}
\end{equation}
where the coefficients $a_i(\tg,q)$ are:
\begin{eqnarray}
a_0 &=& 1                                                              \nonumber\\
a_1 &=& 18\cg(1+q^2)                                                   \nonumber\\
a_2 &=& (1+q^4)(74+49\cg^2)-40q^2+334\cg^2q^2                          \nonumber\\
a_3 &=& 12\cg(1+q^2) \Bigl[ (1+q^4)(29+4\cg^2)+50q^2+64\cg^2q^2 \Bigr] \nonumber\\
a_4 &=& 16\cg^4(1-q^2)^2 (1+34q^2+q^4)
        +2\cg^2(7+58q^2+7q^4)(43+22q^2+43q^4)                          \nonumber\\
    & & -3(1+380q^2-2058q^4+380q^6+q^8)                                \nonumber\\
a_5 &=& 6\cg(1+q^2) \Bigl[ 4\cg^2(1-q^2)^2(19+358q^2+19q^4)
        -1-500q^2+4890q^4-500q^6-q^8 \Bigr]                            \nonumber\\
a_6 &=& 128\cg^4(1-q^2)^4(1+34q^2+q^4)
        - 3\cg^2(1-q^2)^2(1+644q^2-6906q^4+644q^6+q^8)                 \nonumber\\
    & & + 36 q^2(1+34q^2+q^4)^2
\label{eq:a}
\end{eqnarray}
with $\cg=\cos 2\tg$. For any set of parameters $(\g,\tg,q)$, $F>0$
means that the caustic does not have swallowtails, while $F<0$
means that the caustic does have swallowtails. Thus each pair of
solutions to $F=0$ gives an envelope bounding a region in parameter
space in which swallowtails occur.

For $\tg=0^\circ$ and $90^\circ$ we have $\cg=+1$ and $-1$ and the
envelope equation simplifies to
\begin{equation}
  F(\g,\tg,q) = \Bigl[ 1-q^2\pm\g(5+ q^2) \Bigr]^3\,
                \Bigl[ 1-q^2\pm\g(1+5q^2) \Bigr]^3 = 0\,.
\end{equation}
With $\cg=+1$ there are no solutions, and hence no models with
swallowtails. With $\cg=-1$ the envelope of swallowtail models is
\begin{equation}
  {1-q^2 \over 5+q^2} < \g < {1-q^2 \over 1+5q^2}\,.
  \label{eq:t90}
\end{equation}

More generally, for given values of $\tg$ and $q$ the envelope
equation (\ref{eq:F}) is a 6th order polynomial in $\g$ whose roots
are easy to find numerically. Before using a root finder, however,
we should understand how many roots to expect and what their
general ranges are. This requires that we examine how the
polynomial $F$ depends on location in the $(q,\tg)$ plane. First
consider how the number of roots changes as $q$ varies. Additional
pairs of roots appear when $F=0$ and $\partial F / \partial q = 0$,
which can be combined using the resultant method to eliminate $\g$
and obtain the condition
\begin{eqnarray}
  G(\cg,q) &\equiv& 32(1-q^2)^2(1+970q^2+q^4)\cg^4 - (7-986q^2+7q^4)^2 \nonumber\\
  && + \Bigl[ 17(1+q^8)+60196(q^2+q^6)+824358q^4 \Bigr] \cg^2 \quad =\quad 0,
  \label{eq:G}
\end{eqnarray}
where again $\cg = \cos 2\tg$. This is a second order polynomial in
$\cg^2$, so its solution is
\begin{eqnarray}
  \cg^2 = {1 \over 64(1-q^2)^2(1+970q^2+q^4)} &\Biggl[ &
    17(1+q^8)+60196(q^2+q^6)+824358q^4 \nonumber\\
  && +81(1+q^2)\sqrt{1+322q^2+q^4}\quad \Biggr]\ .
  \label{eq:qroots}
\end{eqnarray}
(The second solution of the quadratic equation is unphysical
because it has $c_\g^2<0$.) The right-hand side of \refeq{qroots}
is between 0 and 1 for all $0 \le q \le 1$, so there is always a
physical solution; in fact, there are two solutions, one for
$\cg>0$ and one for $\cg<0$. The two solutions intersect when
$\cg=0$, which occurs at $q=\sqrt{7/(493+90\sqrt{30})}=0.0843$.
Next, consider the behavior of the polynomial $F$ for large $\g$,
which is controlled by the coefficient $a_6$. If $a_6>0$ then $F>0$
for large $\g$, so there are no swallowtails for large $\g$.
However, if $a_6<0$ then $F<0$ for large $\g$, so there are
swallowtails for arbitrarily large $\g$. Thus the behavior of the
polynomial changes qualitatively when $a_6$ changes sign. Since
$a_6$ is a quadratic polynomial in $\cg^2$ (see eq.~\ref{eq:a}), it
is easy to find that $a_6=0$ if
\begin{eqnarray}
  \cg^2 = {3 \over 256(1-q^2)^2(1+34q^2+q^4)} &\Biggl[ &
    (1+644q^2-6906q^4+644q^6+q^8) \nonumber\\
  && \pm(1-253q^2-253q^4+q^6)\sqrt{1-254q^2+q^4}\quad \Biggr]\ .
  \label{eq:a6}
\end{eqnarray}
The right-hand side is real for $|q| \le 8-3\sqrt{7} = 0.0627$, and
at this point it has $\cg=\pm1/\sqrt{28}$ or $\tg=39.55^\circ$ and
$50.45^\circ$.

With these results we can understand the $(q,\tg)$ plane as shown
in Figure 12. The curves given by \refeqs{qroots}{a6} define nine
different regions where the swallowtail envelopes (the parameter
regions with $F<0$) have the following properties:
\begin{eqnarray}
  \mbox{none} &:& \mbox{no envelopes                                                         } \nonumber\\
  1+          &:& \mbox{one envelope, with $\g>0$                                            } \nonumber\\
  2+          &:& \mbox{two envelopes, both with $\g>0$                                      } \nonumber\\
  1+,1-       &:& \mbox{two envelopes, one with $\g>0$ and one with $\g<0$                   } \nonumber\\
  1+,\infty   &:& \mbox{one finite envelope with $\g>0$, and envelopes with $\g\to\pm\infty$ } \nonumber\\
  1-          &:& \mbox{one envelope, with $\g<0$                                            } \nonumber\\
  2-          &:& \mbox{two envelopes, both with $\g<0$                                      } \nonumber\\
  1-,1+       &:& \mbox{two envelopes, one with $\g<0$ and one with $\g>0$                   } \nonumber\\
  1-,\infty   &:& \mbox{one finite envelope with $\g<0$, and envelopes with $\g\to\pm\infty$ } \nonumber
\end{eqnarray}
Negative shear is unphysical (see \S 2.2), so we are interested
only in envelopes with $\g>0$, which occur in six of the nine
regions: $(1+)$; $(2+)$; $(1+,1-)$; $(1+,\infty)$; $(1-,1+)$; and
$(1-,\infty)$.

With this detailed knowledge of the $(q,\tg)$ plane, we can use a
numerical root finder with \refeq{F} to obtain the envelope of
swallowtail models. The results are shown and discussed in \S 3.3.
In particular, Figures 4--6 show envelopes in the $(q,\g)$ plane
for different values of $\tg$. Each envelope corresponds to a
particular horizontal line in Figure 11. Each point where this line
crosses a curve in the $(q,\tg)$ plane corresponds to a cusp in the
envelope in the $(q,\g)$ plane.

\section{Models that can produce 8 images}

We can also ask what range of swallowtail models produce
overlapping swallowtails that bound 8-image regions (e.g.\ Figure
3b). We cannot answer this question analytically for arbitrary
values of the shear angle $\tg$. However, from the examples in \S
3.2 we expect that overlapping swallowtails occur only when the
shear is nearly orthogonal to the galaxy, and the case with
$\tg=90^\circ$ can be studied analytically. When $\tg=90^\circ$ the
system has reflection symmetry, so the points on the caustic with
$\t=0$ and $\t=\pi/2$ are always cusps, no matter how convoluted
the rest of the curve is. (These cusps are indicated in Figure 3.)
Moreover, the $\t=0$ cusp always opens to the left, and the
$\t=\pi/2$ cusp always opens downward. Label the $\t=0$ cusp
position $(u_H,0)$ and the $\t=\pi/2$ cusp position $(0,v_V)$ (H
for horizontal, V for vertical). From Figure 3 we see that there
are overlapping swallowtails only if $u_H$ and $v_V$ are both
positive. (If one is positive and one negative then there are
swallowtails that do not overlap.) We can use \refeq{astr} to
rewrite the conditions $u_H > 0$ as follows:
\begin{eqnarray}
  u_H>0 \ \Longleftrightarrow\ \g<\g_H(q) &\equiv& {1-E_H(q) \over 1+E_H(q)}\,,
  \label{eq:gH} \\
  \mbox{where}\quad E_H(q) &\equiv&
    {q \over \sqrt{1-q^2}}\,\tan^{-1}\left({\sqrt{1-q^2} \over q}\right)\,.
    \nonumber
\end{eqnarray}
We can similarly rewrite the condition $v_V > 0$:
\begin{eqnarray}
  v_V>0 \ \Longleftrightarrow\ \g>\g_V(q) &\equiv& {E_V(q)-1 \over E_V(q)+1}\,,
  \label{eq:gV} \\
  \mbox{where}\quad E_V(q) &\equiv&
    {1 \over \sqrt{1-q^2}}\,\tanh^{-1}\left(\sqrt{1-q^2}\right)\,.
    \nonumber
\end{eqnarray}
For any galaxy axis ratio $0 \le q \le 1$, these functions satisfy
$\g_V(q) \le \g_H(q)$. Thus the condition for overlapping
swallowtails is $\g_V(q) < \g < \g_H(q)$, and this result is shown
in Figure 4.

\newpage

\begin{figure}
\centerline{\epsfysize=6.5in \epsfbox{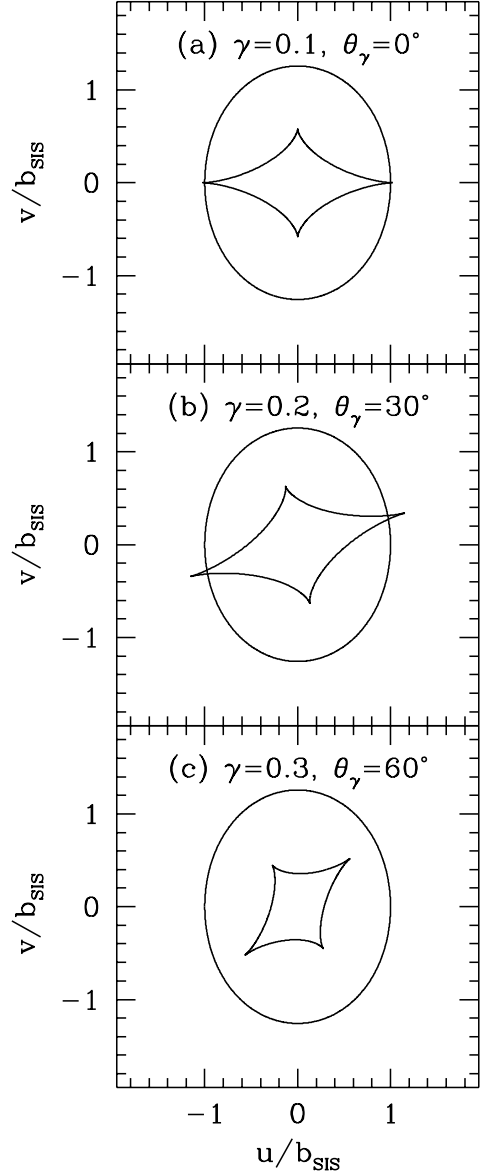}}
\caption{Sample caustics and pseudo-caustics for models with a
singular isothermal ellipsoid galaxy and an external shear. The
smooth elongated curves are the pseudo-caustics and the
diamond-shaped curves are the astroid caustics. All panels have a
galaxy with a projected axis ratio $q=0.5$ and a shear whose
magnitude $\g$ and direction $\tg$ are indicated. The axes are
labeled in terms of the natural lensing length scale $b_{\rm SIS}$
from \refeq{bSIS}.}
\end{figure}

\begin{figure}
\centerline{\epsfysize=6.0in \epsfbox{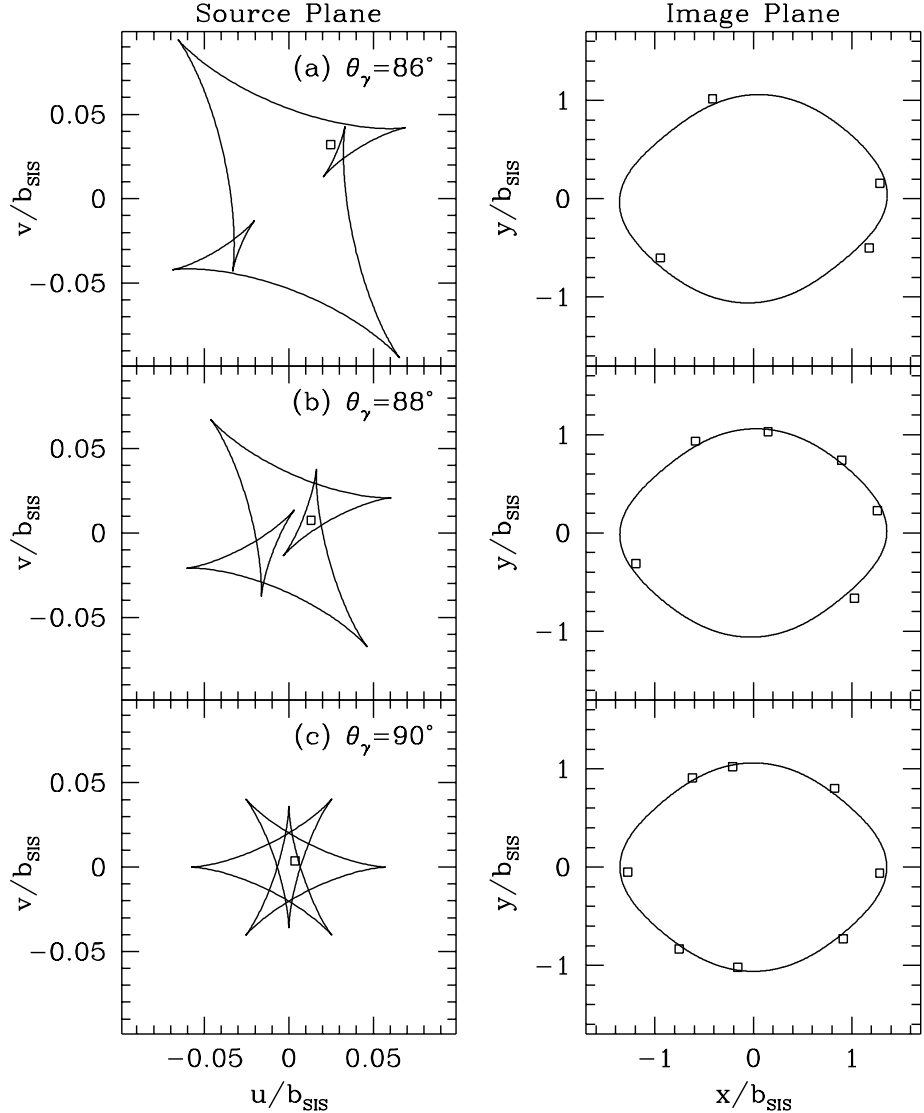}}
\caption{Examples of ``swallowtail'' caustics produced when the
ellipticity and shear are nearly orthogonal. The left-hand panels
show the tangential caustics and sample source positions; the
pseudo-caustics are larger than the frames. The right-hand panels
show the corresponding tangential critical curves and image
positions. All models have $q=0.5$ and $\g=0.22$ and the specified
shear direction $\tg$. The points show that a source inside the
astroid but outside the swallowtails produces 4 images (case a);
a source inside a swallowtail produces 6 images (case b); and a
source inside overlapping swallowtails produces 8 images (case c).
The total magnification of the sample images in cases a, b, and c
is 76.8, 178.3, and 266.2, respectively.}
\end{figure}

\begin{figure}
\centerline{\epsfysize=6.5in \epsfbox{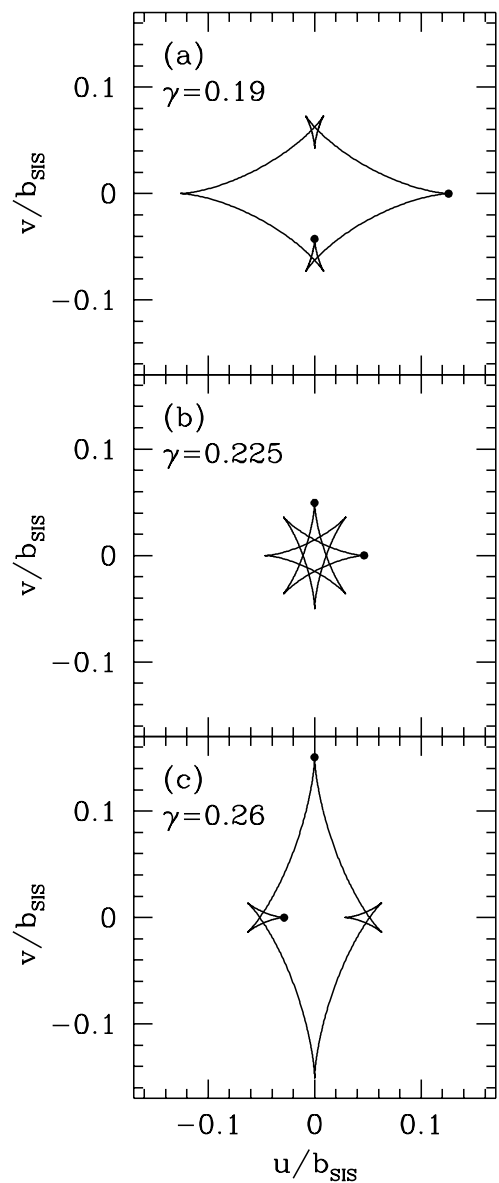}}
\caption{The dependence of swallowtails on the shear. All models
have $q=0.5$ and an orthogonal shear ($\tg=90^\circ$) with the
specified magnitude $\g$. The points indicate the cusps with $\t=0$
(on the horizontal axis) and $\t=\pi/2$ (on the vertical axis); see
Appendix B for details.}
\end{figure}

\begin{figure}
\centerline{\epsfxsize=6.0in \epsfbox{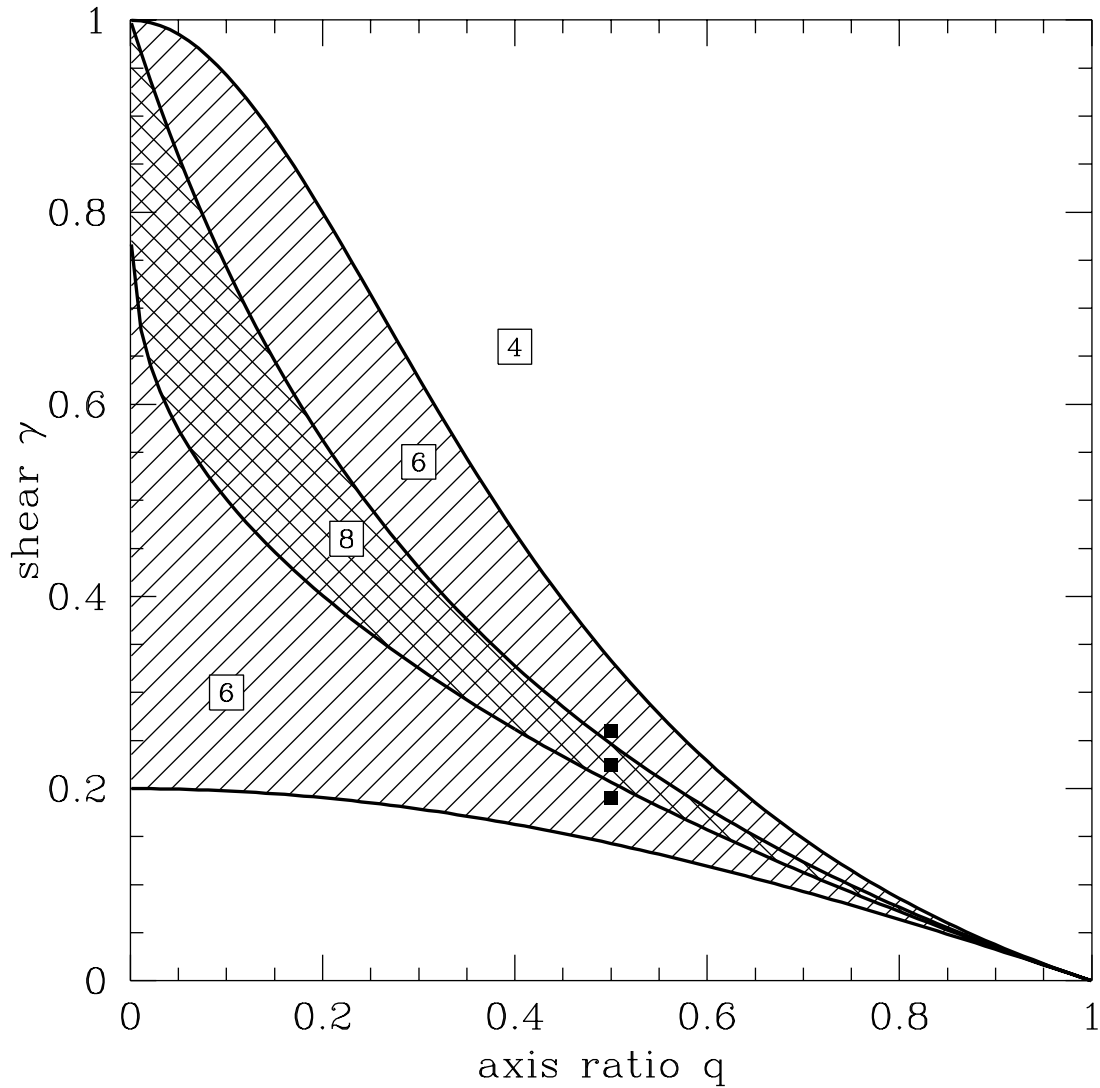}}
\caption{The envelope of galaxy+shear models that can produce at
least 6 images in the $(q,\g)$ plane, for a shear orthogonal to the
lens galaxy ($\tg=90^\circ$). Models in the cross-hatched region
have overlapping swallowtails and can produce up to 8 images.
Models in the shaded region have non-overlapping swallowtails and
can produce up to 6 images. Models outside the shaded region do not
have swallowtails and can produce at most 4 images. The curves
bounding these regions are given by eqs.~(\ref{eq:t90}), (\ref{eq:gH}),
and (\ref{eq:gV}) in Appendices A and B. The three filled points
indicate the locations of the three sample models shown in Figure 3.}
\end{figure}

\begin{figure}
\centerline{\epsfxsize=6.0in \epsfbox{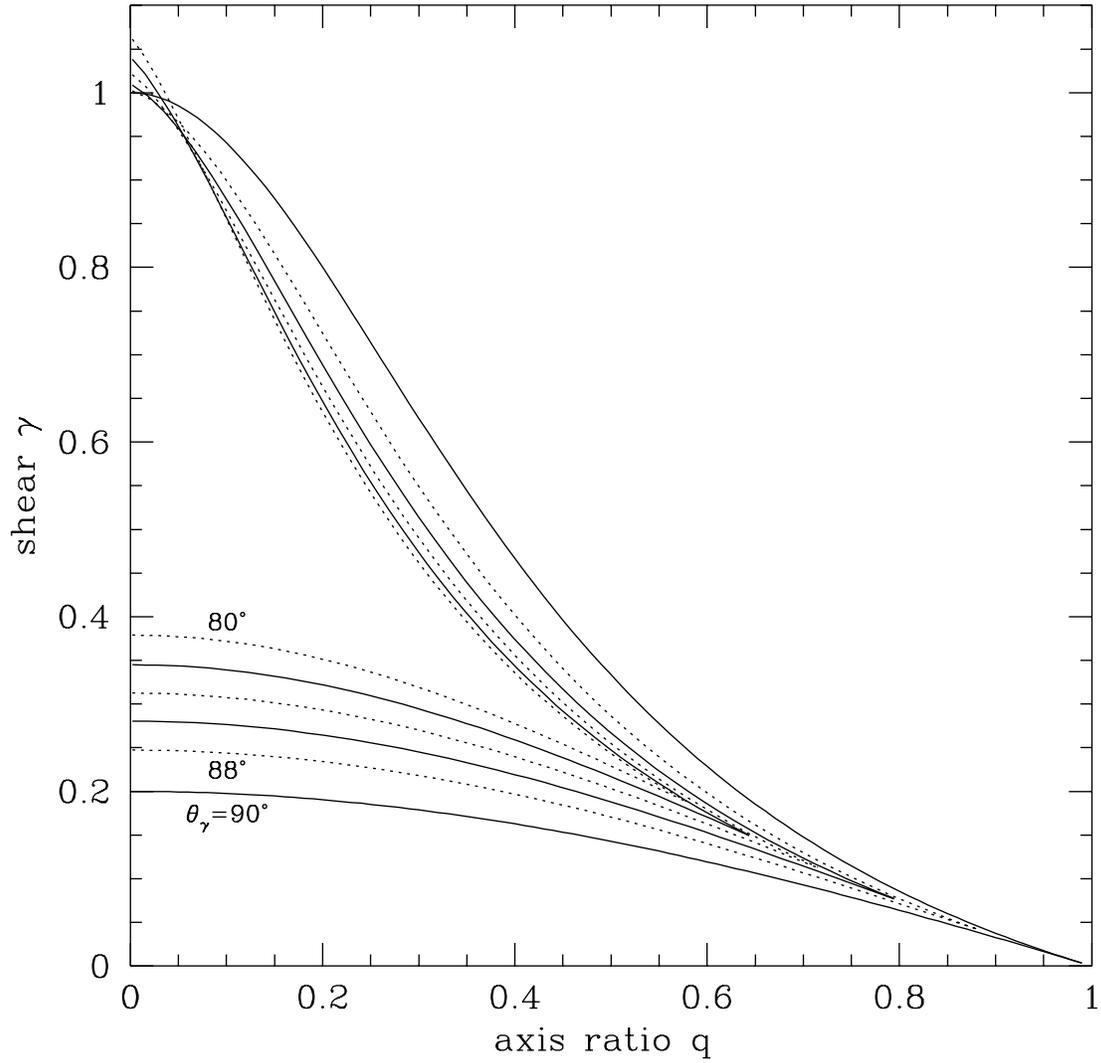}}
\caption{Similar to Figure 4, but for different values of the shear
angle near $\tg \simeq 90^\circ$. Only the swallowtail envelope is
shown (not the envelope for overlapping swallowtails). The outer
envelope corresponds to $\tg=90^\circ$, and moving inward the line
type alternates as $\tg$ decreases by $2^\circ$.}
\end{figure}

\begin{figure}
\centerline{\epsfxsize=6.0in \epsfbox{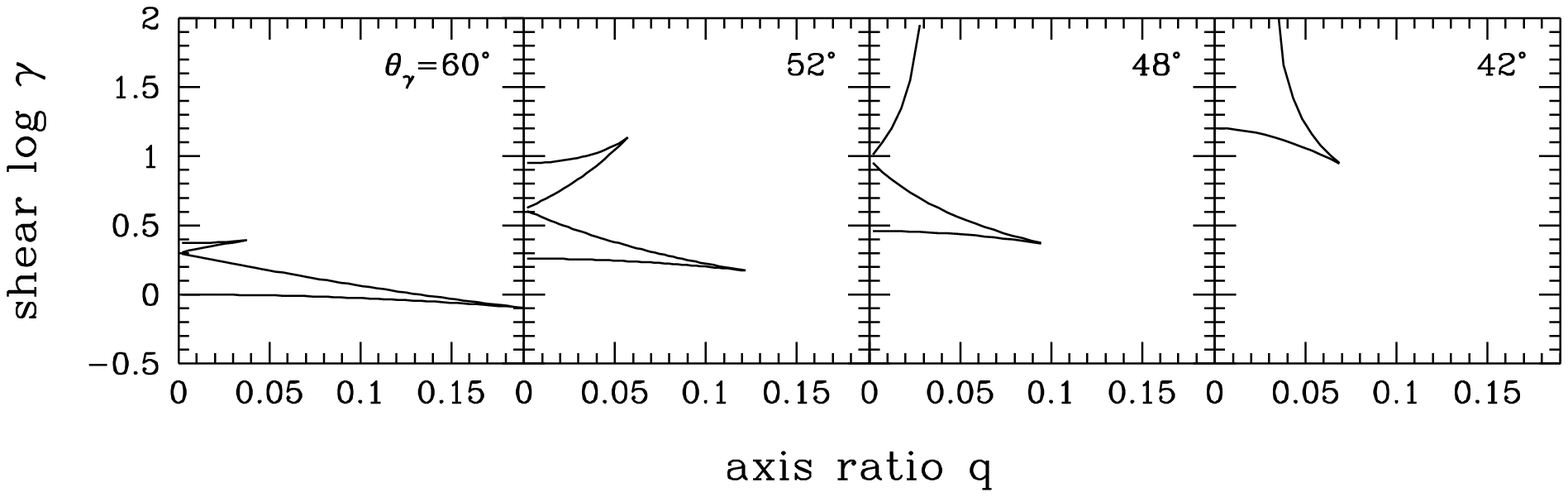}}
\caption{Similar to Figure 4, but for values of the shear angle
near $\tg \simeq 45^\circ$. Note that the vertical axes are
$\log\g$ instead of $\g$. For $\tg > 50.45^\circ$ there are two
closed envelopes. For $50.45^\circ > \tg > 45^\circ$ there is one
closed envelope and one envelope that extends to $\g \to +\infty$.
For $45^\circ > \tg > 39.55^\circ$ there is only the envelope that
extends to infinity. For $\tg < 39.55^\circ$ there is no envelope.
(Also see Figure 12 in Appendix A.)}
\end{figure}

\begin{figure}
\centerline{\epsfxsize=6.0in \epsfbox{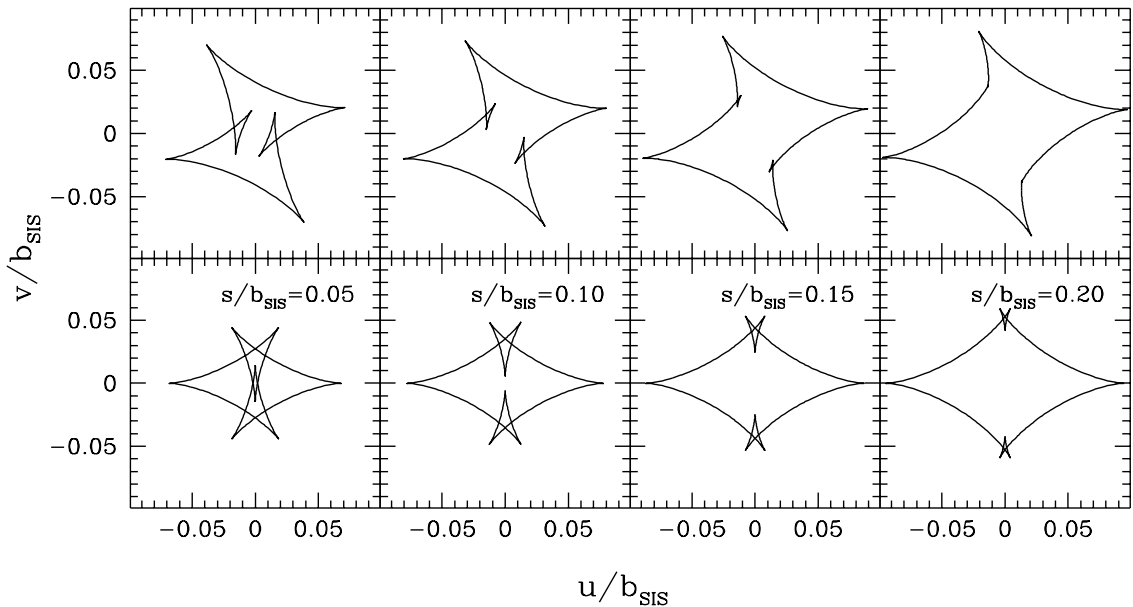}}
\caption{The dependence of swallowtails on the galaxy core radius
$s$. All models have $q=0.5$ and $\g=0.22$. The top panels show
models with $\tg=88^\circ$, while the bottom panels show models
with $\tg=90^\circ$. For comparison, Figures 2b and 2c show the
same models in the limit $s=0$.}
\end{figure}

\begin{figure}
\centerline{\epsfxsize=6.5in \epsfbox{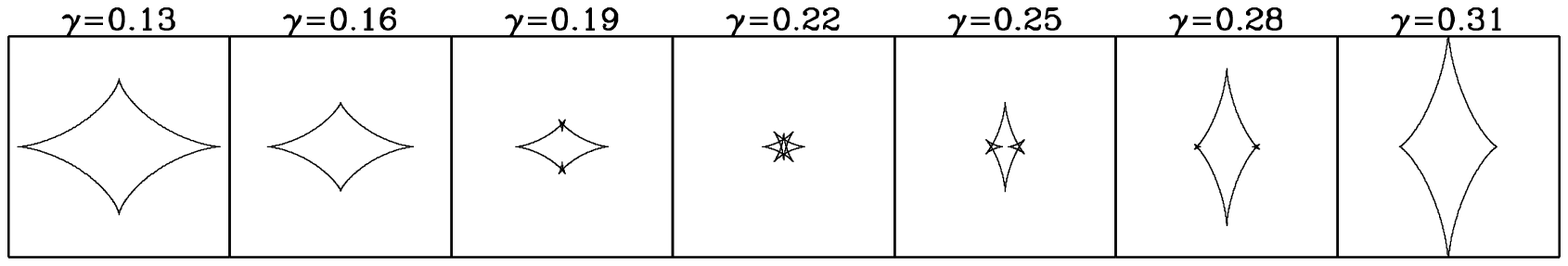}}
\centerline{\epsfxsize=6.5in \epsfbox{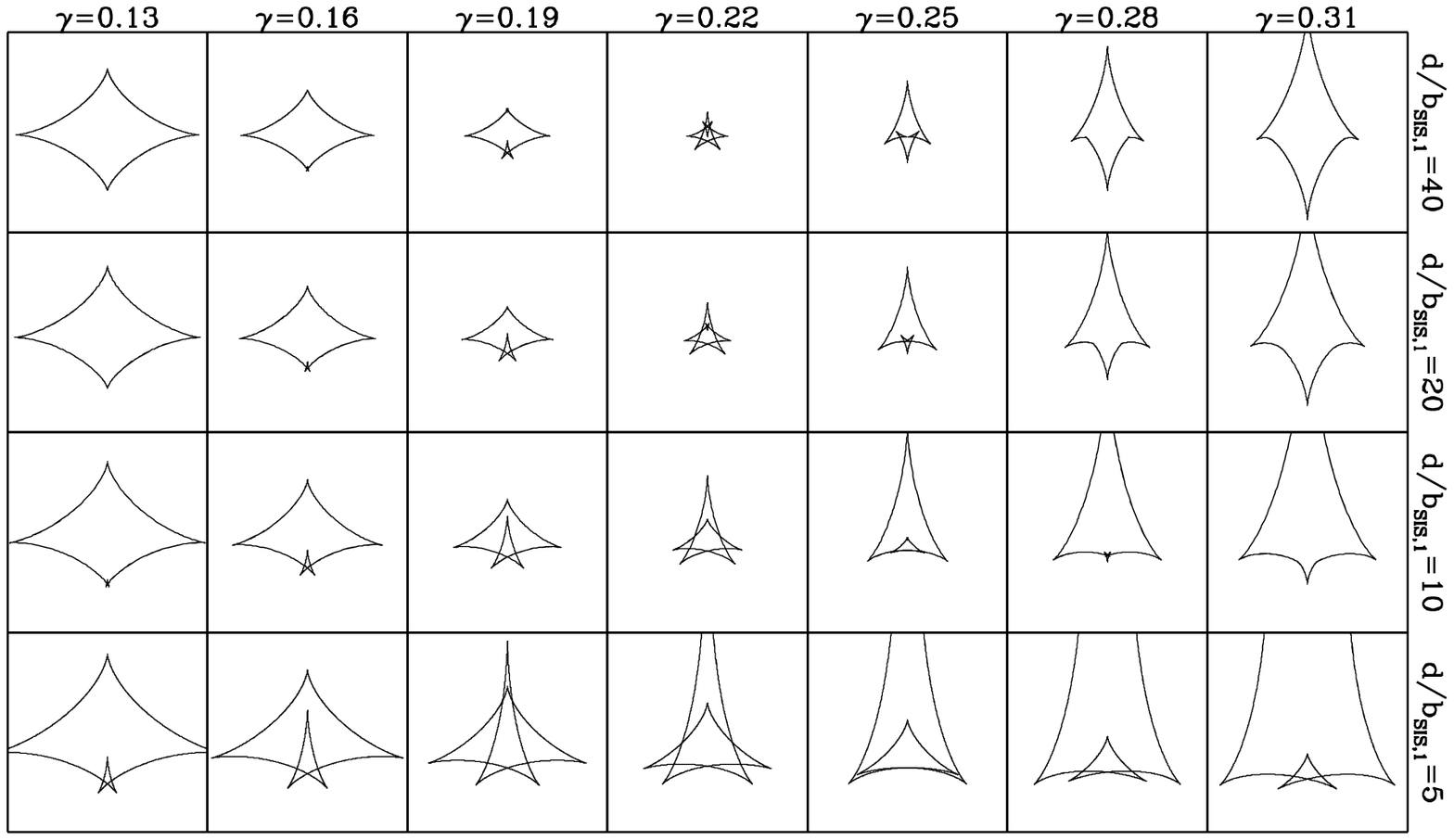}}
\caption{The caustics for a galaxy with $q=0.5$ perturbed by an
external shear (top panels), and by a singular isothermal sphere at
angle $\tg=90^\circ$ and distance $d$ (bottom panels). The strength
of the perturbation is given by $\g$. Each frame is $0.6\,b_{\rm
SIS,1}$ on a side. In some cases with large $\g$ and small $d$, the
caustics of the main lens galaxy and the perturber merge and become
larger than the frames.}
\end{figure}

\begin{figure}
\centerline{\epsfxsize=6.5in \epsfbox{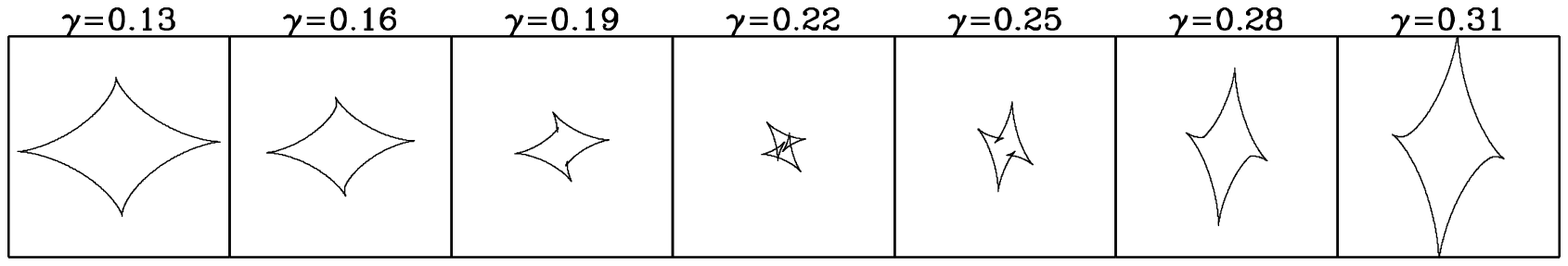}}
\centerline{\epsfxsize=6.5in \epsfbox{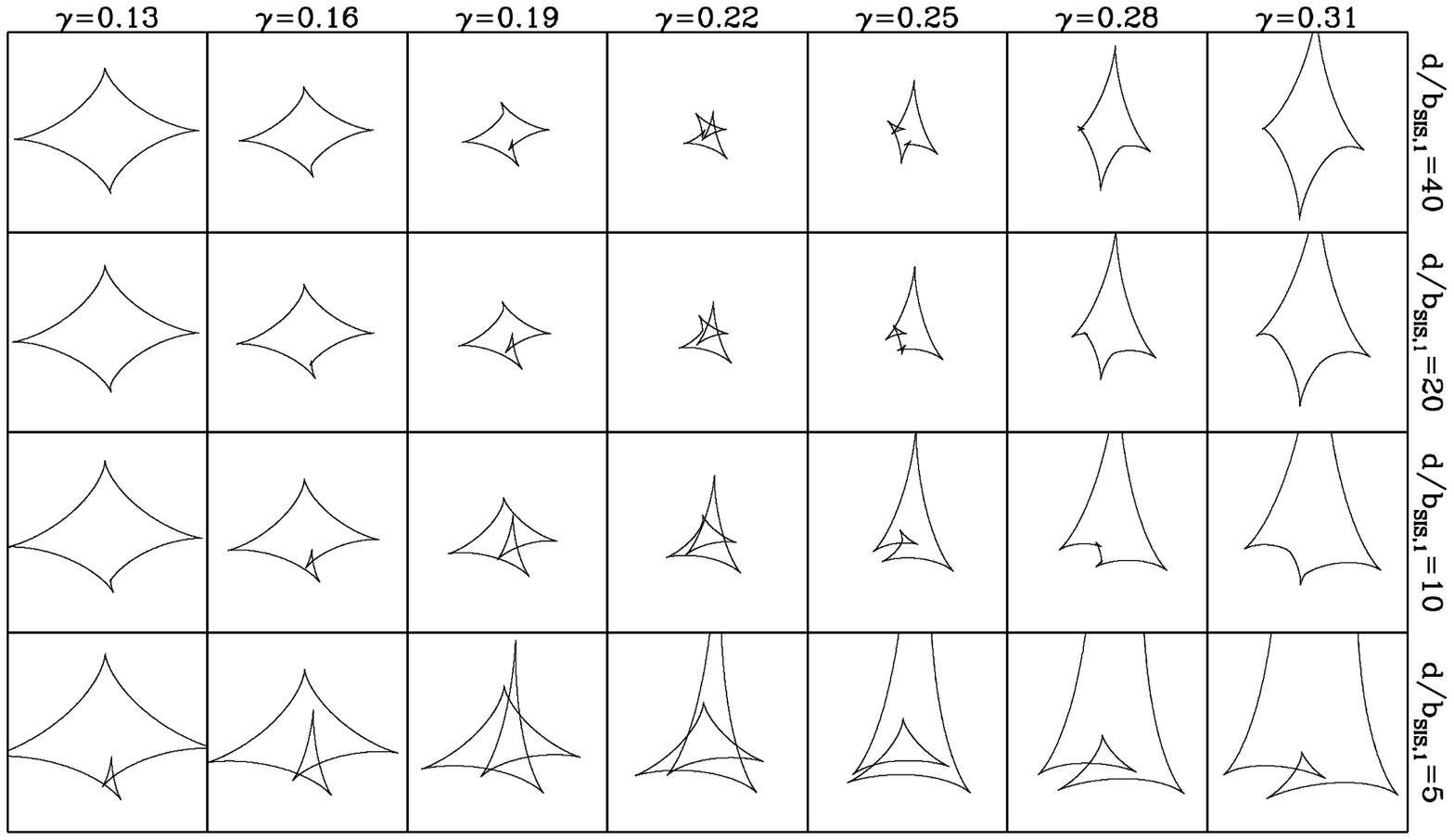}}
\caption{Similar to Figure 8, but for $\tg=88^\circ$. 
Each frame is $0.6\,b_{\rm SIS,1}$ on a side.
}
\end{figure}

\begin{figure}
\centerline{\epsfxsize=5.0in \epsfbox{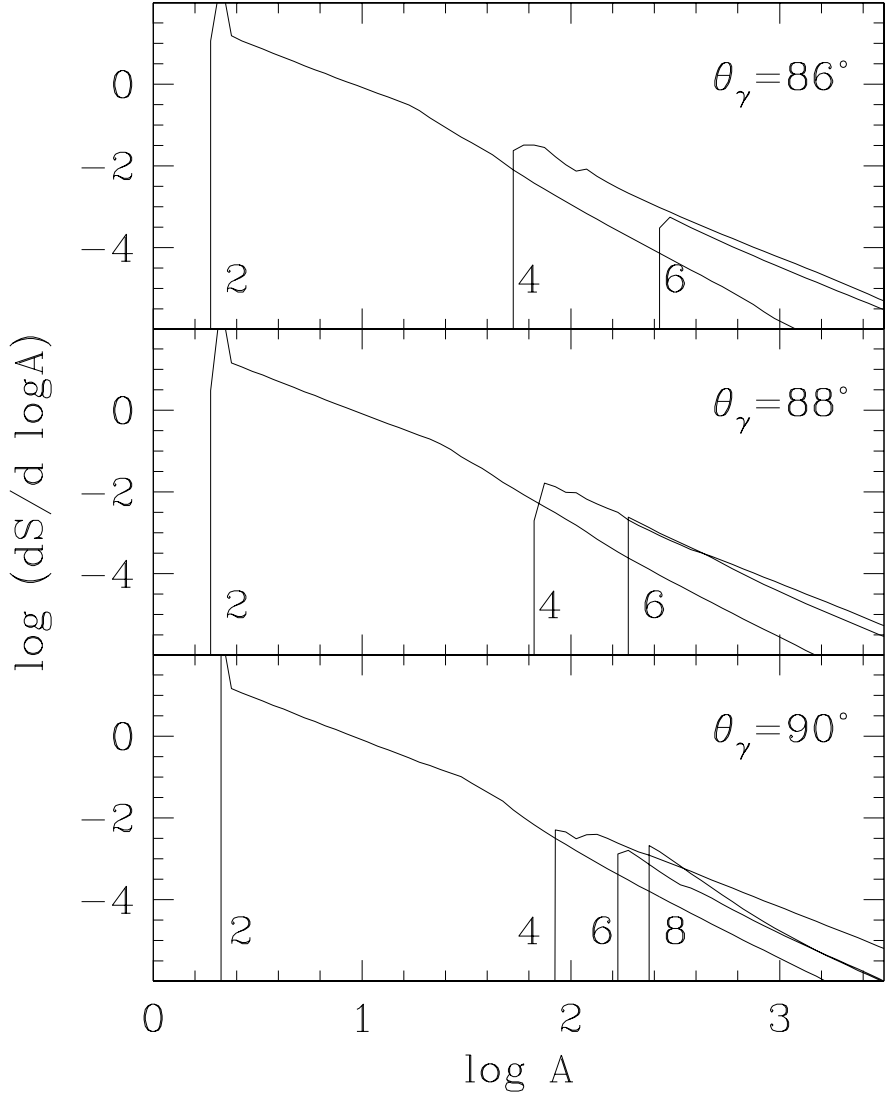}}
\caption{Differential cross sections (in units of $b_{\rm SIS}^2$)
vs.\ logarithm of magnification $A$ for 2-, 4-, 6-, and 8-image
lenses produced by galaxy+shear lens models. The cross-sections are
computed for a point source in a square with side-length of
$5\,\bSIS$. The lens galaxy axis ratio is $q=0.5$, the shear
amplitude is $\g=0.22$, and the shear direction is indicated; the
cases correspond to those shown in Figure 2.}
\end{figure}

\begin{figure}
\centerline{\epsfxsize=6.0in \epsfbox{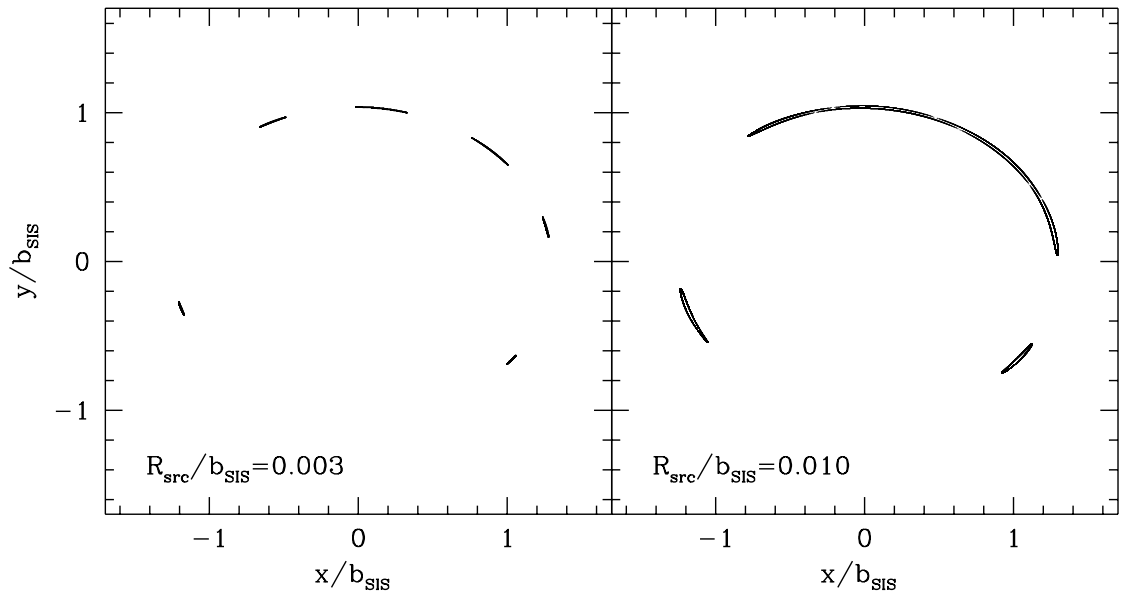}}
\caption{Sample images for a source with finite radius $R_{\rm
src}$. The lens model and source position are the same as in Figure
2b. The images are distorted tangentially relative to the lens
galaxy, and as the source gets bigger they smear out into an
Einstein ring.}
\end{figure}

\begin{figure}
\centerline{\epsfxsize=6.0in \epsfbox{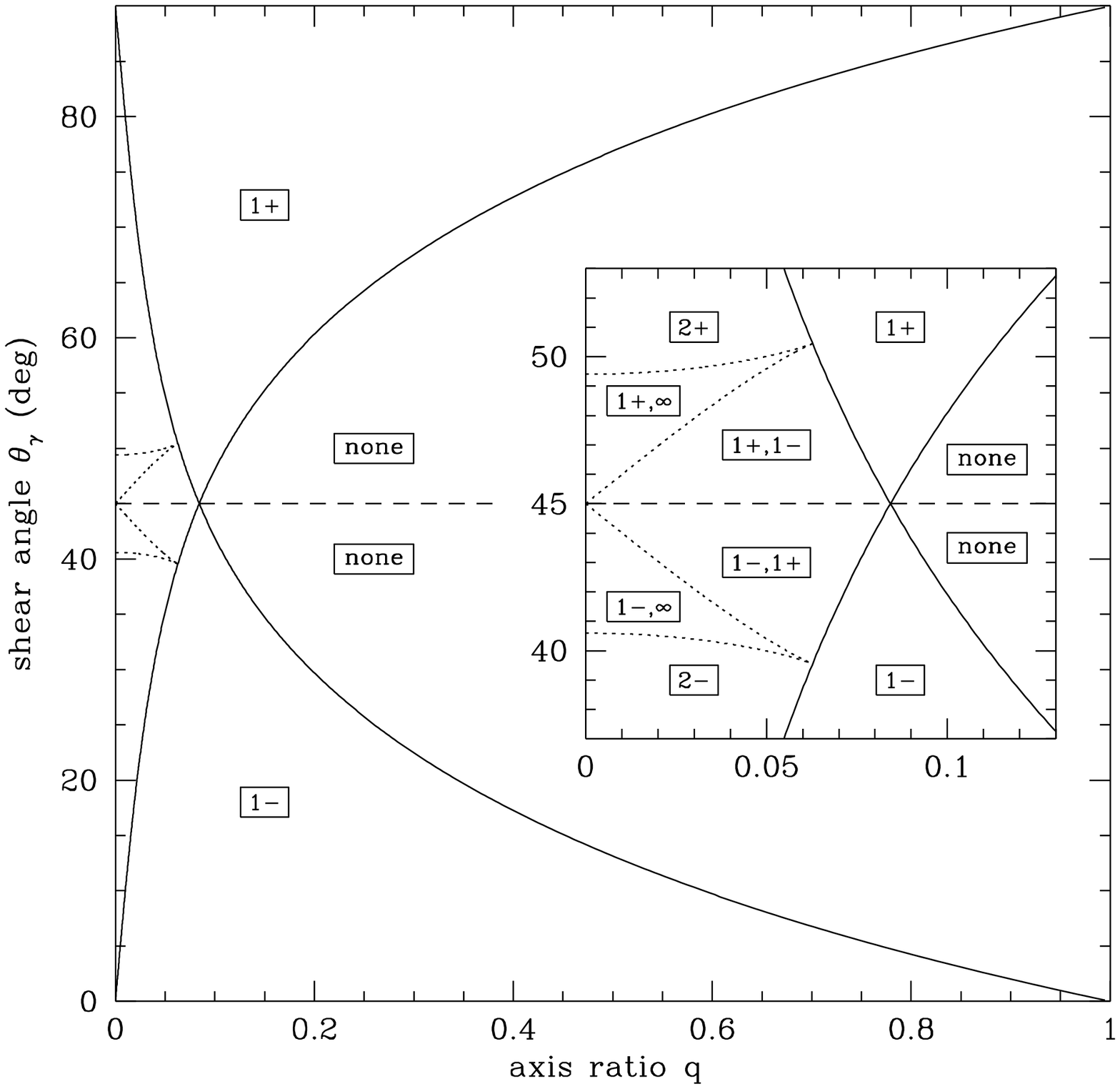}}
\caption{The $(q,\tg)$ plane for galaxy+shear models. The inset
shows a closeup of the region with $q$ small and $\tg \simeq
45^\circ$. The solid curves show solutions to \refeq{qroots},
while the dotted curves show solutions to \refeq{a6}. The solid
curves intersect at $\tg=45^\circ$ and
$q=\radical"270370{7/(493+90\radical"270370{30})}=0.0843$.
The dotted curves meet the solid curves at
$q=8-3\radical"270370{7}=0.0627$ and $\cg=\pm1/\radical"270370{28}$
or $\tg=39.55^\circ$ and $50.45^\circ$. The curves divide the plane
into nine regions. Each region has a distinct set of envelopes
containing swallowtail models, as explained in Appendix A.}
\end{figure}

\end{document}